\documentclass[prd,superscriptaddress,amsfonts,amssymb,amsmath,showpacs,twocolumn]{revtex4-2}
\usepackage{bm}
\usepackage{amsfonts}
\usepackage{latexsym}
\usepackage{graphicx}
\usepackage{amsmath}
\usepackage{palatino}
\usepackage{mathpazo}
\usepackage{textcomp}
\usepackage{float}
\usepackage{booktabs}
\usepackage{dcolumn}
\usepackage{booktabs}
\usepackage{multirow}
\usepackage{hyperref}
\hypersetup{colorlinks,citecolor=blue}
\usepackage{amsmath}
\usepackage{xcolor}
\usepackage{orcidlink}
\usepackage{epsfig}
\usepackage{caption}
\usepackage{subcaption}
\usepackage{commath}
\hypersetup{colorlinks,citecolor=blue}
\hypersetup{colorlinks=true,linkcolor=magenta,filecolor=magenta,    urlcolor=blue}

\usepackage{amssymb}
\usepackage{relsize,exscale}

\def\be{\begin{equation}}
\def\ee{\end{equation}}
\def\bea{\begin{eqnarray}}
\def\eea{\end{eqnarray}}

\begin{document}

\title{\bf Estimation of \( H_0 \) and \( r_d \) in the \(\omega(z)\) Parameterization within Einstein and Horava-Lifshitz Gravity Using DESI-Y1 and SDSS-IV}

\author{Ujjal Debnath}
\email{ujjaldebnath@gmail.com} 
\affiliation{Department of
Mathematics, Indian Institute of Engineering Science and
Technology, Shibpur, Howrah-711 103, India}
\author{Himanshu Chaudhary}
\email{himanshu.chaudhary@stud.ubbcluj.ro} 
\affiliation{Department of Physics, Babeș-Bolyai University, Kogălniceanu Street, Cluj-Napoca, 400084, Romania,}
\affiliation{Research Center of Astrophysics and Cosmology, Khazar University, Baku, 41 Mehseti Street, AZ1096, Azerbaijan}
\author{Niyaz Uddin Molla}
\email{niyazuddin182@gmail.com}
\affiliation{Department of
Mathematics, Indian Institute of Engineering Science and Technology, Shibpur, Howrah-711 103, India}
\author{S. K. J. Pacif}
\email{shibesh.math@gmail.com}
\affiliation{Pacif Institute of Cosmology and Selfology (PICS), Sagara, Sambalpur 768224, Odisha, India}
\author{G.Mustafa}
\email{gmustafa3828@gmail.com} 
\affiliation{Department of Physics,
Zhejiang Normal University, Jinhua 321004, Peoples Republic of China}

\begin{abstract}
We present a novel dynamical dark energy model within the frameworks of both Einstein gravity and Horava-Lifshitz gravity. Utilizing CDMMA parametrization of the dark energy equation of state \(\omega(z)\), we derive solutions to the field equations. By employing recent cosmological datasets, such as cosmic chronometer datasets, Type Ia Supernovae datasets, and Baryonic Oscillation datasets (DESI Y1 and SDSS-IV). We validate our model and determine optimal parameter values. Furthermore, we analyze the evolution of the Universe by showing the redshift dependence plots of key cosmological parameters through graphical representations. We also perform diagnostic analyses to compare our model with the standard model. Using the Akaike Information Criterion (AIC), we compare the three models and find that all of them are supported by the current data, making it impossible to discard any of them. Our model aligns well with recent observations and unveils intriguing features of the Universe, particularly the late-time behavior of the Universe.  
\end{abstract}
\maketitle
\section{Introduction}\label{sec1}
Modern cosmology began with Albert Einstein's formulation of General Relativity in 1915, transforming our understanding of gravity and the universe's structure. In 1929, Edwin Hubble's observations revealed that galaxies are receding from us, indicating that the universe is expanding and supporting the Big Bang theory. This model was further solidified by the discovery of the cosmic microwave background radiation in 1965 by Arno Penzias and Robert Wilson. Initially, it was believed that the universe's expansion rate was slowing due to gravity. However, in the late 1990s, observations of distant Type Ia supernovae by teams led by Saul Perlmutter, Brian Schmidt, and Adam Riess suggested that the expansion is accelerating \cite{SupernovaSearchTeam:1998fmf,SupernovaCosmologyProject:1998vns,spergel2003first,SupernovaSearchTeam:1998fmf,perlmutter1999measurements}. This unexpected acceleration implied the existence of a mysterious force with high negative pressure, termed dark energy, which now is thought to constitute about two-thirds of the universe's total energy density \cite{Riess:2006fw,Jimenez:2003iv,Peebles:2002gy,Capozziello:2005tf,WMAP:2008ydk,Nojiri:2005pu,li2011dark}. This discovery was a pivotal moment, profoundly impacting our understanding of the cosmos and its fate. Since then, modern cosmology has increasingly relied on astrophysical observations to refine theoretical models. Observations of the cosmic microwave background and supernovae have revealed that the universe has experienced two major periods of rapid expansion: an early phase known as inflation and a later phase of acceleration starting around six billion years after the Big Bang and continuing today. These findings challenge our current understanding and underscore the need for further research into the causes of the universe's accelerated expansion.\\\\
Dark energy is a theoretical energy form thought to pervade the cosmos, exerting a significant negative pressure that fuels the acceleration of cosmic expansion. This mysterious component is classified based on its equation of state parameter, which gives rise to various candidates of dark energy. These candidates offer potential insights into the fundamental forces at play in shaping the universe's trajectory. When the EOS falls within $-1 < \omega < -\frac{1}{3}$, they are referred to as quintessence-type dark energy, and when $\omega < -1$, they are classified as phantom-type dark energy. Other models of dark energy can cross the phantom divide $\omega = -1$ from both sides and are termed quintom-type dark energy \cite{Quintessence1,Quintessence2,Quintessence3,Quintessence4,dark1,dark2,dark3,dark4,dark5,dark6,dark7,frieman2008dark}. Another idea of explaining the late time cosmic acceleration involves the concept of modifying gravity as an alternative to dark energy. This approach introduces models where gravity behaves differently than predicted by classical General Relativity. These modified gravity models emerge from theories such as string/M-theory, offering a gravitational alternative to exotic forms of matter. Unlike dark energy, these models don't require negative kinetic terms. They provide explanations for phenomena like early-time inflation and the smooth transition from deceleration to late-time acceleration in the universe. While these modified gravity theories are gaining attention in contemporary research as rivals to General Relativity, they face rigorous constraints that must be addressed.\\\\
Despite axed by several times, GR stood on a high that explain gravity so well and provides a more comprehensive framework that accurately predicts a wide range of phenomena, from the bending of light by gravity to the expansion of the universe. Its importance in studying cosmology lies in its ability to explain the large-scale structure and dynamics of the universe, including black holes, gravitational waves, and the overall geometry of space and time. General Relativity has been crucial in the development of modern cosmological models and continues to be a foundational theory for understanding the universe's behavior and evolution. However, it fails to explain certain phenomena e.g. the late-time acceleration, initial singularity and a few more. At infra red scale, it needs modification as well as at the quantum scale. Several modifications have been proposed in the past few decades and one of them is Horava-Lifshitz gravity.\\\\
Horava-Lifshitz gravity represents a departure from conventional Einsteinian gravity, offering a fresh perspective on the dynamics of gravity, particularly at the quantum and cosmological scales. Originating from Petr Horava's work in 2009, this theory integrates elements of quantum field theory and renormalization group flow into gravitational physics. Unlike general relativity, it introduces a distinctive scaling relationship between space and time, potentially resolving longstanding issues with the behavior of quantum gravity at high energies. This unique approach has stirred significant interest and discussion among theoretical physicists, prompting further exploration into its implications for our understanding of gravity and the universe's fundamental nature. This framework has generated significant attention but has also faced challenges and criticisms, particularly about its viability as a quantum theory and the presence of ghost instabilities. Researchers are still studying and improving the Horava-Lifshitz gravity framework, particularly its effects on cosmology, black hole physics, and its larger impact on our awareness of basic interactions in the universe. It is crucial to mention that the scientific community continuously argues and discusses the feasibility and consequences of Horava-Lifshitz gravity \cite{Horava1,Horava2,Hovrava3,Horava4,chaudhary2024cosmological,khurana2024exploring,chaudhary2024investigating,maity2024constraining,para5}.\\\\
Numerous cosmological models have emerged in attempts to explain late-time cosmic acceleration. Despite decades of theoretical exploration, a definitive, universally accepted model remains elusive. A recent approach to understanding the accelerating Universe involves parametrizing the equation of state parameter for dark energy (DE) at a phenomenological level. This method aims to explore evolutionary scenarios without bias toward any specific DE model, seeking to uncover the characteristics of the mysterious component driving cosmic acceleration. Referred to as the model-independent approach, it relies on estimating model parameters from existing observational datasets. However, this method faces limitations: some parameterizations encounter challenges related to divergence, while others may not fully capture nuanced insights into the true nature of dark energy due to constraints imposed by the assumed parametric form. Various investigations have been undertaken to elucidate the cosmic acceleration of the Universe by employing viable parameterizations of the Equation of State (EoS) \cite{Debnath:2020rho,Escamilla-Rivera:2019aol,Capozziello:2005pa,para5}.\\\\
Motivated by these EoS parameterizations, researchers have also extensively discussed and scrutinized the parametrization of the deceleration parameter, Hubble parameter, and a few other geometrical and physical parameters in the literature. A brief summary can be found in \cite{pacif1,pacif2,pacif3,pacif4,pacif2017reconstruction,mishra2023cosmological,q(z)1,q(z)2,q(z)3,q(z)4,q(z)5,q(z)6,q(z)7,q(z)8,q(z)9,q(z)10,q(z)11,q(z)12,EoSNew1,EoSnew2,EoSnew3,Eosnew4}. As the Universe evolves, it shifts from a deceleration phase to a late-time acceleration phase \cite{moraes2016transition}. Consequently, any cosmological model must include a transition from deceleration to acceleration to fully explain the Universe's evolution. The deceleration parameter, denoted as \(q = -\frac{a\ddot{a}}{\dot{a}^2}\), where \(a(t)\) represents the customary scale factor, plays a pivotal role in determining whether the Universe is undergoing acceleration (\(q < 0\)) or deceleration (\(q > 0\)).\\\\
In this work, we aim to consider CDMMA parametrization of the equation of state parameter \(\omega(z)\) and solve the Einstein field equations in both Einsteinian gravity and Horava-Lifshitz gravity. The paper is organized as follows: Section \ref{sec1} provides an introduction regarding late-time cosmic acceleration and the motivation for parametrization schemes. In Sections \ref{sec2} and \ref{sec3}, the basic equations of Einstein and Horava-Lifshitz gravity are derived. Section \ref{sec4} discusses the parametrization of \(\omega(z)\) and the solutions to the field equations are derived. In Section \ref{sec5}, we provide comprehensive details about the datasets used and describe the methodology and the dataset used to obtain the best fit values of model parameters. In Section \ref{sec4}, we compare the model predictions with real observational datasets and the standard $\Lambda$CDM model. In Section \ref{sec7}, we thoroughly explore cosmographic parameters, placing a specific emphasis on the deceleration, jerk, and snap parameters. Section \ref{sec8} specifically focuses on conducting the Statefinder and \(O_m\) diagnostics. Finally, Sections \ref{sec9} and \ref{sec10} are reserved for the discussion of results with statistical analyses and the presentation of concluding remarks.
\section{Basic Equations in Einstein's Gravity}\label{sec2}
We view the Einstein-Hilbert action in the following manner.
\begin{equation}\label{EH}
{\cal S}=\int d^4x\sqrt{-g}\left[\frac{R-2\Lambda}{16\pi
G}+{\cal{L}}_m\right]
\end{equation}
In the context, the line element characterizing the non-flat Friedmann-Lemaître-Robertson-Walker-Metric (FLRW) Universe is expressed as follows: where $R$ denotes the Ricci scalar, $\Lambda$ represents the cosmological constant, ${\cal L}_{m}$ stands for the matter Lagrangian, and $g$ signifies the determinant of $g_{ij}$, with the choice of $c=1$. This formulation captures the homogeneous and isotropic nature of the Universe under consideration.
\begin{equation}\label{FRW}
ds^2=-dt^2 +a^2(t)\left[\frac{dr^2}{1-kr^2}+r^2\left(d\theta^2 +
sin^2 \theta d\phi^2 \right)\right]
\end{equation}
where $a(t)$ denotes the scale factor and $k$ stands for the flat, open, and closed models of the Universe, the Einstein's field equation is expressed as $G_{ij}=8\pi G T_{ij}$, where the energy-momentum tensor of the fluid is specified as follows.
\begin{eqnarray}
T_{\mu\nu}=(\rho+p)u_\mu u_\nu+p g_{\mu\nu}
\end{eqnarray}
The quantities $\rho$ and $p$ represent the energy density and pressure density of the fluid, respectively. The fluid's 4-velocity, denoted by $u^\mu=\frac{dx^\mu}{ds}$, obeys the condition $u^\mu u_\mu=-1$. In Einstein's gravity, with the presence of a cosmological constant, the Friedmann equations for the FLRW metric take the following form:
\begin{equation}\label{F1}
H^2+\frac{k}{a^2}=\frac{8\pi G}{3}~\rho+\frac{\Lambda}{3}
\end{equation}
and
\begin{equation}\label{F2}
\dot{H}-\frac{k}{a^2}=-4\pi G(\rho+p)
\end{equation}
In this context, the Hubble parameter as $H$, where it is defined as the derivative of the scale factor $a$ with respect to cosmic time $t$, represented by $\dot{a}/a$. We are examining a Universe filled with a fluid content characterized by energy density $\rho$ and pressure $p$. These quantities adhere to the energy conservation equation.
\begin{equation}\label{Cons}
\dot{\rho}+3H(\rho+p)=0
\end{equation}
Consider Universe with dark matter (DM) and dark energy (DE), total energy density ($\rho$) and pressure ($p$) are expressed as $\rho = \rho_m + \rho_d$ and $p = p_m + p_d$, respectively. With separate conservation equations for DM and DE, we have:
\begin{equation}\label{DM}
    \dot{\rho}_m + 3H(\rho_m + p_m) = 0,
\end{equation}
and
\begin{equation}\label{DE}
    \dot{\rho}_d + 3H(\rho_d + p_d) = 0.
\end{equation}
Dark matter is pressureless, with \( p_m = 0 \), Eq \eqref{DM} leading to \( \rho_m = \rho_{m0}a^{-3} \). The equation of state parameter \( w(z) = p/\rho \) yields \( \rho_d = \rho_{d0}~e^{3\int \frac{1+w(z)}{1+z} dz} \), where \( \rho_{m0} \) and \( \rho_{d0} \) are the present values of the energy densities of dark matter and dark energy respectively. Define the dimensionless density parameters:
\begin{equation*}\label{dp}
\begin{aligned}
\Omega_{m0} &= \frac{8\pi G\rho_{m0}}{3H_{0}^{2}}, \Omega_{d0} = \frac{8\pi G\rho_{d0}}{3H_{0}^{2}}, \Omega_{k0} = -\frac{k}{H_0^2}, \\
~\text{and }\Omega_{\Lambda0} &= \frac{\Lambda}{3H_0^2}.
\end{aligned}
\end{equation*}
So from Eq (\ref{F1}), we obtain the Hubble parameter as
\begin{equation}\label{H}
\begin{aligned}
H^{2}(z) &= H_{0}^{2}\left[\Omega_{m0}(1+z)^{3} + \Omega_{k0}(1+z)^{2}\right. \\
&\quad \left. + \Omega_{\Lambda 0} + \Omega_{d0}~ e^{3\int \frac{(1+w(z))}{1+z}dz}\right]
\end{aligned}
\end{equation}
where $z$ is the redshift parameter described as $1+z=\frac{1}{a}$
(presently, $a_{0}=1$) and
$\Omega_{m0}+\Omega_{k0}+\Omega_{\Lambda 0}+\Omega_{d0}=1$.
\section{Basic Equations in Horava-Lifshitz Gravity}\label{sec3}
It is useful to utilize the Arnowitt-Deser-Misner (ADM) decomposition of the metric as presented in \cite{H1,H2,H3}.
\begin{equation}\label{HL1}
ds^2=-N^2dt^2+g_{i j} \left(dx^i+N^i dt\right)\left(dx^j+N^j dt\right).
\end{equation}
The lapse function \(N\), the shift vector \(N_i\), and the spatial metric \(g_{ij}\) are fundamental components in the formulation of Hořava-Lifshitz (HL) gravity. The coordinates transform according to the scaling transformation \(t \rightarrow l^3 t\) and \(x^i \rightarrow l x^i\). The HL gravity action consists of two main parts: the kinetic term and the potential term, expressed as
\begin{equation}
S_g = S_k + S_v = \int dt \, d^3x \, \sqrt{g} \, N \left( L_k + L_v \right),
\end{equation}
where the kinetic term is defined by
\[ S_k = \int dt \, d^3x \, \sqrt{g} \, N \left[ \frac{2 \left( K_{ij}K^{ij} - \lambda K^2 \right)}{\kappa^2} \right], \]
with \(\kappa^2 = 32\pi G\). The extrinsic curvature \(K_{ij}\) is given by
\begin{equation}
K_{ij} = \frac{\dot{g}_{ij} - \nabla_i N_j - \nabla_j N_i}{2N}.
\end{equation}
The Lagrangian \(L_v\) for the potential term contains several invariants that can be simplified using its symmetry properties \cite{H4,H5,H6}, which are derived from the principle of detailed balance. Taking detailed balance into account, the expanded form of the action can be expressed as follows:
\begin{equation}
S_g = \int dt \, d^3x \, \sqrt{g} \, N \left[ \frac{2 \left( K_{ij}K^{ij} - \lambda K^2 \right)}{\kappa^2} + L_v \right].
\end{equation}
This formulation ensures that the action respects the desired symmetries and scaling properties essential to Hořava-Lifshitz gravity.
\begin{equation}
\begin{aligned}
S_g &= \int dt d^3x \sqrt{g} N \Bigg[\frac{2\left(K_{ij}K^{ij} - \lambda K^2\right)}{\kappa^2} + \frac{\kappa^2 C_{ij}C^{ij}}{2\omega^4} - \\
& \frac{\kappa^2 \mu \epsilon^{i j k } R_{i, j} \Delta_j R^l_k}{2\omega^2 \sqrt{g}} \quad + \frac{\kappa^2 \mu^2 R_{ij} R^{ij}}{8} \\
& - \frac{\kappa^2 \mu^2}{8(3\lambda-1)}\left\{\frac{(1-4\lambda)R^2}{4} +\Lambda R -3 \Lambda^2 \right\}\Bigg],
\end{aligned}
\end{equation}
The Cotton tensor and all covariant derivatives are determined with respect to the spatial metric \( g_{ij} \). Here, \( \epsilon^{ijk} \) is the totally antisymmetric unit tensor, \( \lambda \) is a dimensionless constant, and \( \kappa \), \( \omega \), and \( \mu \) are constants. Assuming the lapse function depends only on time (i.e., \( N \equiv N(t) \)), Horava derived a gravitational action. Using the FLRW metric with \( N = 1 \), \( g_{ij} = a^2(t)\gamma_{ij} \), and \( N^i = 0 \), where
\[ \gamma_{ij} dx^i dx^j = \frac{dr^2}{1 - kr^2} + r^2 d\Omega_2^2, \]
with \( k = -1, 1, 0 \) representing an open, closed, and flat universe respectively, and by varying \( N \) and \( g_{ij} \), we obtain the Friedmann equations.
\begin{equation}\label{HLFriedmann1}
\begin{aligned}
H^2 &= \frac{\kappa^2\rho}{6\left(3\lambda-1\right)} \\
&\quad + \frac{\kappa^2}{6\left(3\lambda-1\right)}\left[\frac{3\kappa^2\mu^2 k^2}{8\left(3\lambda-1\right)a^4} + \frac{3\kappa^2\mu^2 \Lambda^2}{8\left(3\lambda-1\right)}\right] \\
&\quad - \frac{\kappa^4 \mu^2 \Lambda k}{8\left(3\lambda-1\right)^2a^2}.
\end{aligned}
\end{equation}
\begin{equation}\label{HLFriedmann2}
\begin{aligned}
\dot{H} + \frac{3H^2}{2} &= -\frac{\kappa^2 p}{4(3\lambda-1)} \\
&\quad - \frac{\kappa^2}{4(3\lambda-1)}\left[\frac{3\kappa^2\mu^2 k^2}{8(3\lambda-1)a^4} + \frac{3\kappa^2\mu^2 \Lambda^2}{8(3\lambda-1)}\right] \\
&\quad - \frac{\kappa^4 \mu^2 \Lambda k}{8(3\lambda-1)^2a^2}.
\end{aligned}
\end{equation}
The term proportional to \(\frac{1}{a^4}\) represents a distinctive feature of Hořava-Lifshitz (HL) gravity and can be interpreted as a "dark radiation term"  . Additionally, the constant term \(\Lambda\) corresponds to the cosmological constant. Let's define \(\nu\) as \(\frac{\kappa^{4} \mu^{2} \Lambda}{8(3\lambda -1)^{2}}\). Notably, when \(\lambda=1\), Lorentz invariance is preserved. However, for \(\lambda \ne 1\), Lorentz invariance is violated. Consequently, we can rewrite the above Friedmann equations as follows:
\begin{equation}\label{H1}
H^2=\frac{8\pi G_c}{3}\left(\rho_m +
\rho_{d}\right)+\left(\frac{k^2\nu} {2\Lambda
a^4}+\frac{\nu\Lambda}{2}\right)-\frac{k\nu}{a^2},
\end{equation}
\begin{equation}\label{H2}
\dot{H}+\frac{3}{2}H^2=-4\pi G_c p_d -\left(\frac{k^2\nu}{4\Lambda
a^4}+\frac{3\nu\Lambda}{4}\right)-\frac{k\nu}{2a^2}.
\end{equation}
Using the dimensionless parameters (\ref{dp}) and field equation
(\ref{H1}), we obtain
\begin{equation}\label{E}
\begin{aligned}
H^2(z) &= H_0^2 \Bigg[\frac{2\Omega_{m0}}{3\lambda-1}(1+z)^3 + \nu\Omega_{k0}(1+z)^2 \\
&\quad + \frac{3\nu\Omega_{\Lambda 0}}{2} + \frac{\nu\Omega_{k0}^2(1+z)^4}{6\Omega_{\Lambda 0}} \\
&\quad + \frac{2\Omega_{d0}}{3\lambda-1}~e^{3\int \frac{1+w(z)}{1+z} dz}\Bigg]
\end{aligned}
\end{equation}
with
\begin{equation}
\frac{2}{3\lambda-1}(\Omega_{m0}+\Omega_{d0})+\nu\Omega_{k0}+\frac{3\nu\Omega_{\Lambda
0}}{2}+\frac{\nu\Omega_{k0}^2}{6\Omega_{\Lambda 0}}=1
\end{equation}
The observational data analysis for linear, CPL and JBP models in HL gravity have been studied in \cite{H7}.
\section{CDMMA Parametrization}\label{sec4}
With the preceding discussions on Einstein gravity (EG) and Horava-Lifshitz (HL) gravity, coupled with our impetus to explore the parametrization of cosmological parameters, we introduce a novel proposition: CDMMA parametrization for the equation of state of dark energy. Our proposed formulation aims to encapsulate the nuanced behavior of dark energy within the cosmic landscape. By delineating the equation of state in a specific form, we aspire to enhance our comprehension of the intricate interplay between dark energy, gravity, and the late time behaviour of the universe. We consider here the parametrization of EoS as given by,
\begin{equation}
w(z)=\alpha+\frac{m~w_{1}(1+z)^m+n~w_{2}(1+z)^{n}}{w_{0}+w_{1}(1+z)^m+w_{2}(1+z)^{n}}
\end{equation}
where $\alpha,m,n,w_0,w_1,w_2$ are constants (better call them model parameters). For
$\alpha=-1,m=1,n=2$, we recover the ``ASSS"
(Alam-Sahni-Saini-Starobinski) parametrization \cite{H8,H9}.
For $w_0=1,w_1=0,n=1$, we recover the ```PADE-I" parametrization
\cite{H10,H11,H12}. For $w_1=0$, we can get $``CDMMA"$
(Chaudhary-Debnath-Mustafa-Maurya-Atamurotov)-type parametrization
\cite{H13}. In this case, we obtain the solution to the field equations as,
\begin{equation}
\rho_{d}=\rho_{d0} (1+z)^{3(1+ \alpha)} \left[w_0+w_1 (1+z)^m+w_2
(1+z)^n\right]^3
\end{equation}
and,
{\bf Einstein's gravity:}  From Eq (\ref{H}), we obtain
\begin{equation}
\begin{aligned}
H^2(z) = & H_0^2\left[\Omega_{m0}(1+z)^3+\Omega_{k0}(1+z)^2 +\Omega_{\Lambda0}\right. \\
& \left. + \left(1-\Omega_{m0}-\Omega_{k0}-\Omega_{\Lambda0}\right) \right. \\
& \left. \quad \quad \quad \quad (1+z)^{3(1+ \alpha)} \left[\begin{aligned}w_0&+w_1 (1+z)^m  \\
&+w_2 (1+z)^n\end{aligned}\right]^3\right]
\end{aligned}
\end{equation}
{\bf Horava-Lifshitz gravity:}  From Eq (\ref{E}), we obtain
\begin{equation}
\begin{aligned}
H^2(z) = & H_0^2\left[\frac{2\Omega_{m0}}{3\lambda-1}(1+z)^3+\nu\Omega_{k0}(1+z)^2 \right. \\
& +\frac{3\nu\Omega_{\Lambda 0}}{2}+\frac{\nu\Omega_{k0}^2(1+z)^4}{6\Omega_{\Lambda 0}} \\
& \left. +\left(1-\frac{2}{3\lambda-1}\Omega_{m0}-\nu\Omega_{k0}-\frac{3\nu\Omega_{\Lambda 0}}{2}-\frac{\nu\Omega_{k0}^2}{6\Omega_{\Lambda 0}}\right) \right. \\
& \left. \quad \quad \quad \quad (1+z)^{3(1+ \alpha)} \left[\begin{aligned}w_0&+w_1 (1+z)^m  \\
&+w_2 (1+z)^n\end{aligned}\right]^3\right]
\end{aligned}
\end{equation}
With the solution in hand, we now explore the behavior of the Universe both physically and geometrically. However, to proceed effectively, we need to find the approximate values of the model parameters involved in our model. We must estimate the parameters for which, we use the following methodology.\\\\ 
\section{Methodology}\label{sec5}
In cosmology, parameter estimation typically employs a Bayesian framework to compute the posterior distribution of parameters \(\theta\) given observed data \(D\):
\begin{equation}
P(\theta \mid D) = \frac{\mathcal{L}(D \mid \theta) P(\theta)}{P(D)}.
\end{equation}
Here, \(\mathcal{L}(D \mid \theta)\) represents the likelihood function, \(P(\theta)\) is the prior distribution of the parameters, and \(P(D)\) is the marginal likelihood. Bayesian parameter estimation explores the parameter space \(\theta\), often using algorithms like Metropolis-Hastings \cite{akeret2013cosmohammer}, which guides a random walker through the space, favoring regions of higher likelihood. The mean and error of each parameter are typically determined by examining where the walker spends most time and its deviation in the parameter space. In cases where the posterior distribution is approximately Gaussian, information criteria provide simpler insights into model preference \cite{bernardo2022parametric}. In this study, we investigate CDMMA parameterizations for the dark energy equation of state (EoS) in both Einstein’s gravity and Horava-Lifshitz gravity frameworks, estimating optimal values for the free parameters by maximizing the likelihood function \(\mathcal{L}\) or, equivalently, by minimizing the chi-square statistic, \(\chi^2\). To efficiently explore the parameter space and obtain reliable parameter estimates, we perform a Markov Chain Monte Carlo (MCMC) analysis using the \textbf{emcee} package \cite{foreman2013emcee}. For visualization and plotting of posterior distributions, we utilize the \textbf{GetDist} package \cite{lewis2019getdist}, which provides a robust framework for interpreting and presenting parameter constraints.
\subsection{Cosmological data sets}
\subsubsection{Cosmic chronometers (CC)}
The cosmic chronometers serve as reliable probes of cosmic expansion, offering a model-independent approach to measuring the Hubble constant and the rate of expansion. By studying the age and metallicity of nearby passive galaxies, we can estimate the expansion rate $H_{\mathrm{CC}}(z)$ at a redshift $z_{\mathrm{CC}}$. This approximation is derived from the observation that the expansion rate can be approximated as the ratio of redshift difference to time difference, adjusted by the redshift itself: $H_{\mathrm{CC}}(z) \approx -(\Delta z_{\mathrm{CC}} / \Delta t) /(1+z_{\mathrm{CC}})$. We gather cosmic chronometer observations from various sources \cite{CC1,CC2,CC3,CC4,CC5,CC6} spanning the redshift range $0.07 \lesssim z \lesssim 1.97$, as compiled in \cite{bernardo2022parametric}. These observations provide direct constraints on the expansion history of the universe. To assess the agreement between theoretical predictions and cosmic chronometer measurements at different redshifts, we compute the chi-squared distance $\chi_{\mathrm{CC}}^{2}$:
\begin{equation}
\chi^2_{CC}(\theta) = \Delta H^T(z) \mathbf{C}^{-1} \Delta H(z),
\end{equation}
The term \( \Delta H(z) \) represents the difference between the model's predicted expansion rate, \( H_{\mathrm{Model}}(z) \), and the observed cosmic chronometer data, \( H_{\mathrm{Data}}(z) \) at specific redshift and \( \mathbf{C} \) is the covariance matrix.
\subsubsection{Type Ia supernova (SNIa)}
The Pantheon+ dataset comprises light curves for 1701 Type Ia Supernovae (SNe Ia) from 1550 distinct events, covering a redshift range \(0 \leq z \leq 2.3\) \cite{brout2022pantheon}. The primary observable for SNe Ia is the apparent magnitude:
\begin{equation}m(z) = 5 \log_{10} \left( \frac{d_L (z)}{\text{Mpc}} \right) + \mathcal{M} + 25,
\end{equation}
where \(\mathcal{M}\) represents the absolute magnitude of SNe Ia. In a spatially flat Friedmann-Lemaître-Robertson-Walker (FLRW) Universe, the luminosity distance \(d_L\) is expressed as \cite{brout2022pantheon}:
\begin{equation}
d_L (z) = c(1 + z) \int_{0}^{z} \frac{dz'}{H(z')},
\end{equation}
where \(c\) is the speed of light in units of km/s. The chi-square (\(\chi^2\)) statistic for SNe Ia data is calculated using
\begin{equation}
\chi^2_{\text{SNe Ia}} = \Delta \mathbf{D}^T \mathbf{C}^{-1}_{\text{total}} \Delta \mathbf{D},
\end{equation}
where the total covariance matrix, \(\mathbf{C}_{\text{total}} = \mathbf{C}_{\text{stat}} + \mathbf{C}_{\text{sys}}\), is the sum of statistical and systematic covariance matrices. The vector \(\Delta \mathbf{D}\) of SNe Ia distance moduli deviations is given by: $\Delta \mathbf{D} = \mu(z_i) - \mu_{\text{model}} (z_i, \theta),$ where \(\mu(z_i) = m(z_i) - \mathcal{M}\) represents the observed distance modulus of each SNe Ia. We marginalize the likelihood \( L \propto e^{-\chi^2/2} \) over \( M \), effectively projecting out \( M \) up to a normalization constant \cite{escobal2021cosmological}.
\subsubsection{Baryon acoustic oscillations (BAO)}
\begin{figure*}
\centering
\includegraphics[scale=0.7]{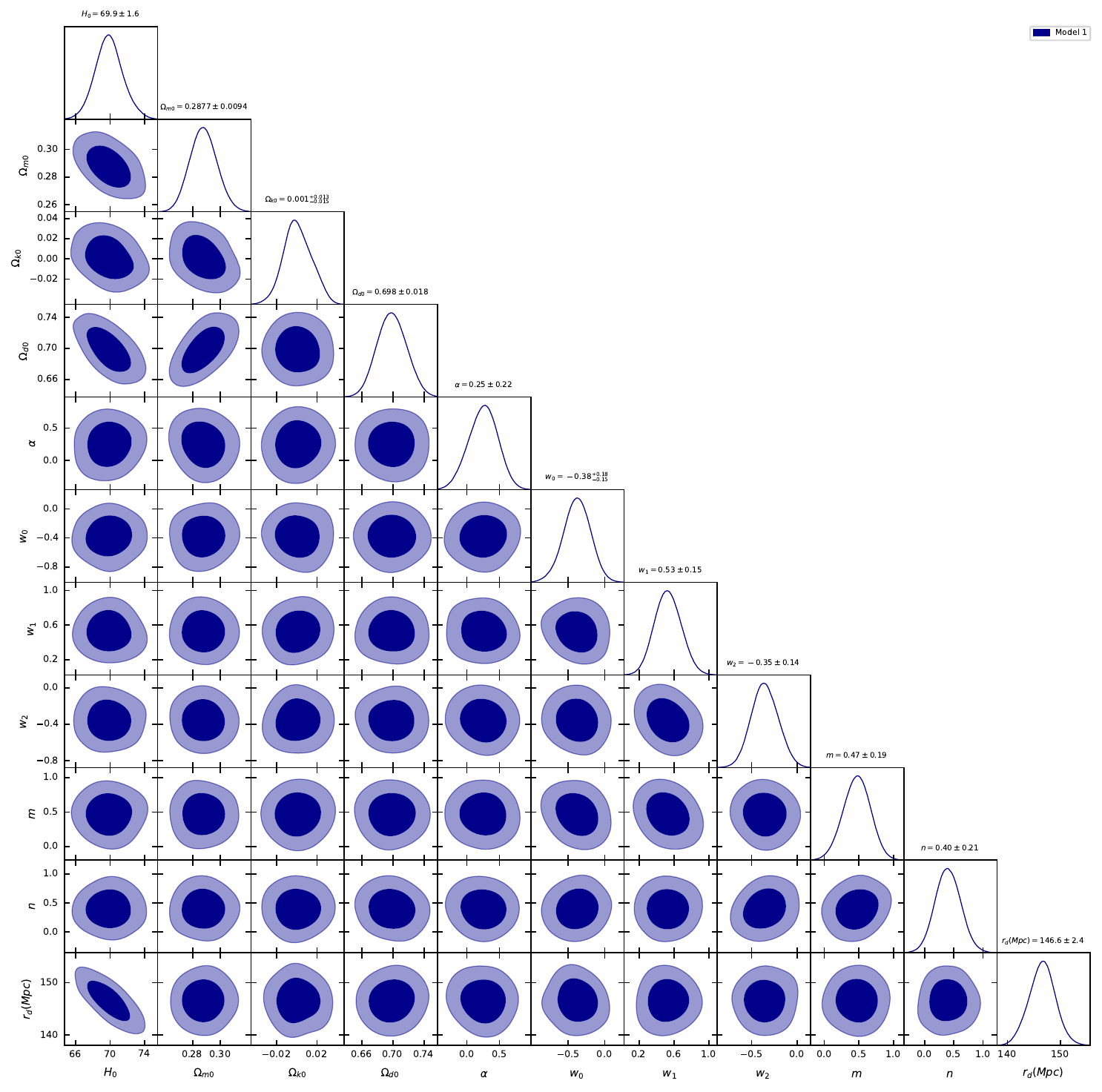}
\caption{The figure shows the posterior distributions of the parameters for the CDMMA parameterization in Einstein's gravity}\label{fig_1}
\end{figure*}

\begin{figure*}
\centering
\includegraphics[scale=0.6]{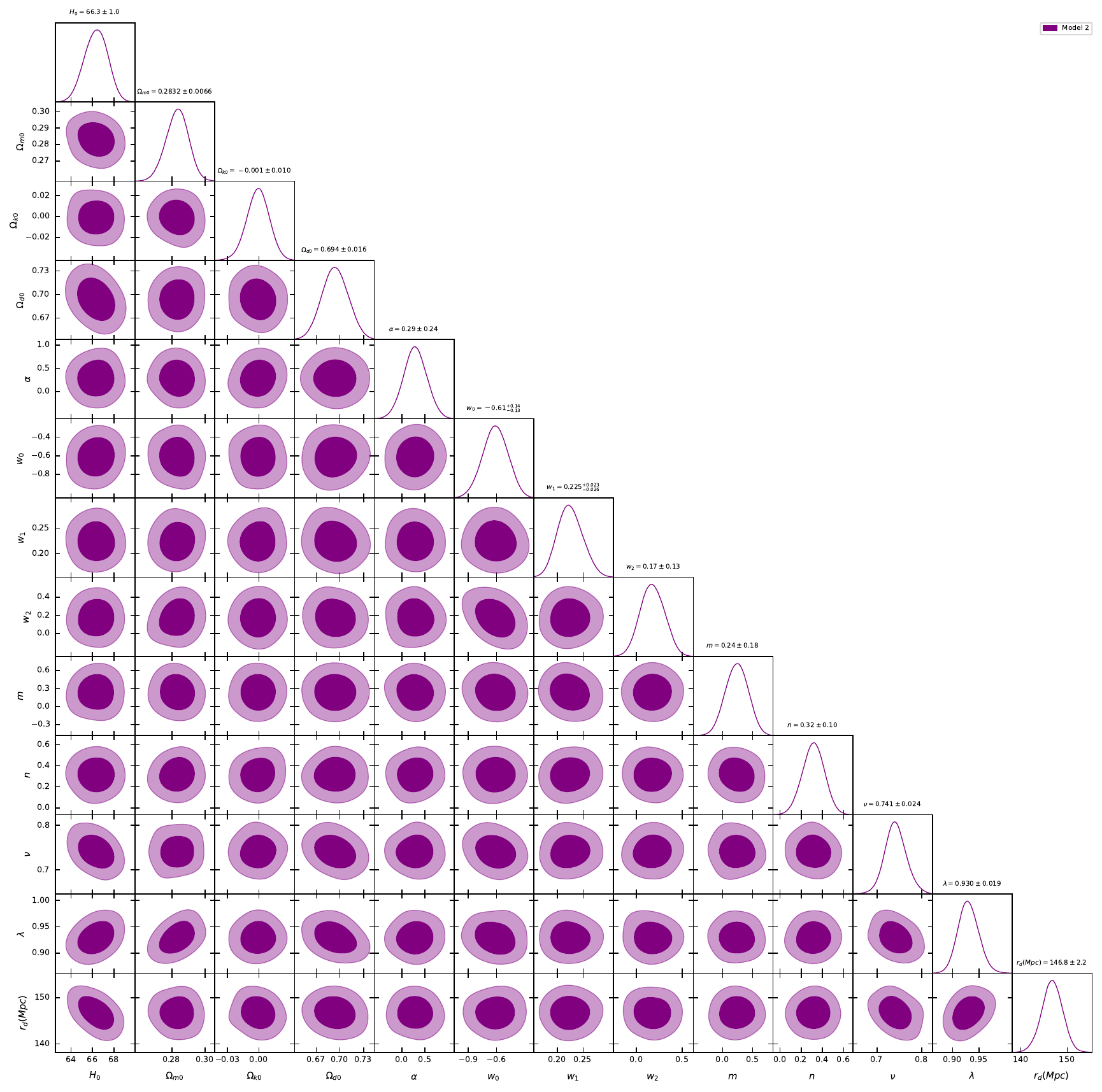}
\caption{The figure shows the posterior distributions of the parameters for the CDMMA parameterization in Horava-Lifshitz gravity}\label{fig_2}
\end{figure*}

In  cosmology, Baryon Acoustic Oscillation (BAO) measurements rely on the sound horizon, \( r_d \), at the baryon decoupling epoch (\( z_d \approx 1060 \)). This sound horizon is given by:
\[
r_d = \frac{1}{H_0} \int_{z_d}^{\infty} \frac{c_s(z)}{E(z)} \, dz,
\]
where \( c_s(z) \) is the sound speed, dependent on the baryon-to-photon density ratio, and \( E(z) \) \(E(z)\) is the dimensionless Hubble parameter. The sound horizon \( r_d \) serves as a calibration scale for BAO observations, typically set with a prior from CMB Planck measurements. In this analysis, however, we exclude the CMB-derived prior for \( r_d \) and treat \( r_d \) as a free parameter \cite{jedamzik2021reducing,pogosian2024consistency,lin2021early,vagnozzi2023seven}. In this analysis, we use BAO datasets from the first-year observations of the Dark Energy Spectroscopic Instrument (DESI Y1) \cite{adame2024desi}, along with data from the completed Sloan Digital Sky Survey (SDSS-IV) \cite{alam2021completed}. BAO measurements in the transverse direction provide an estimate of \(\frac{D_H(z)}{r_d} = \frac{c}{H(z) r_d}\), where \(D_H(z)\) is the Hubble distance at redshift \(z\). We also use the comoving angular diameter distance \(\frac{D_M(z)}{r_d}\) in a flat Universe, where $D_M(z) = \frac{c}{H_0} \int_0^z \frac{dz'}{E(z')}.$ Additionally, we incorporate the volume-averaged distance \(D_V(z) / r_d\), which encodes the BAO peak position. This distance is defined as $D_V(z) = \left[z D_H(z) D_M^2(z)\right]^{1/3}.$ To compare theoretical predictions with observational data from BAO measurements, we calculate the distance function \(\chi_{\mathrm{BAO}}^{2}\) as follows: The \(\chi^2\) for the distance measurements scaled by the sound horizon, \( D_X / r_d \), is given by:
\begin{equation}
\chi^2_{D_X / r_d} = \Delta D_X^T \cdot \mathbf{C}^{-1}_{D_X} \cdot \Delta D_X,
\end{equation}
where \( \Delta D_X = D_{X / r_d, \text{Model}} - D_{X / r_d, \text{Data}} \) for \( X = H, M, V \), and \( \mathbf{C}_{D_X}^{-1} \) is the inverse covariance matrix for each \( X \). For more details, refer to \cite{chaudhary2024semi}. Figs \ref{fig_1} and \ref{fig_2} display the confidence contours for each parameter of the CDMMA parameterization within both the EG and HL gravity frameworks. Table \ref{tab_1} presents the best-fit values for each parameter in these frameworks.
\begin{table}
\begin{tabular}{|c|c|c|c|}
\hline
Models & Parameter & Prior & JOINT  \\
\hline
& $H_{0}$ & $[50.,100.]$ & $68.2{\pm 1.1}$  \\[0.1cm]
$\Lambda$CDM Model & $\Omega_{m0}$ &$[0,1.]$   & $0.3119{\pm 0.0085}$  \\[0.1cm]
& $\Omega_{d0}$ &$[0,1.]$   & $0.6866{\pm 0.0078}$  \\[0.1cm]
&$r_{d} (Mpc)$   &$[100,300]$  &$146.6{\pm 2.4}$  \\[0.1cm]
\hline
\multirow{12.6}{*}{Model 1} 
& $H_{0}$ &$[50.,90.]$   &$69.9{\pm 1.6}$ \\[0.1cm]
& $\Omega_{mo}$ &$[0.,1.]$  &$0.2877{\pm 0.0094}$\\[0.1cm]
& $\Omega_{ko}$ &$[-0.1,0.1]$   &$0.001_{-0.015}^{+0.013}$ \\[0.1cm]
& $\Omega_{d0}$ &$[0,1.]$   & $0.698{\pm 0.018}$  \\[0.1cm]
& $\alpha$ &$[0.,1]$   &$0.25{\pm 0.22}$ \\[0.1cm]
& $w_{0}$ &$[-1.,0]$   &$-0.38_{-0.15}^{+0.18}$ \\[0.1cm]
& $w_{1}$ &$[0.,1]$   &$0.53{\pm 0.15}$ \\[0.1cm]
& $w_{2}$ &$[-1.,0]$   &$-0.35{\pm 0.15}$ \\[0.1cm]
& $m$  &$[0,1.]$   & $0.47{\pm 0.19}$  \\[0.1cm]
& $n$  &$[0,1.]$   & $0.40{\pm 0.21}$  \\[0.1cm]
&$r_{d}$ (Mpc)   &$[100,300]$  &$146.6{\pm 2.4}$  \\[0.1cm]
\hline
\multirow{12.6}{*}{Model 2} 
& $H_{0}$ &$[50.,90.]$   &$66.3{\pm 1.0}$ \\[0.1cm]
& $\Omega_{mo}$ &$[0.,1.]$  &$0.2832{\pm 0.0066}$ \\[0.1cm]
& $\Omega_{ko}$ &$[-0.1,0.1]$   &$-0.001{\pm 0.010}$ \\[0.1cm]
& $\Omega_{d0}$ &$[0,1.]$   & $0.694{\pm 0.016}$  \\[0.1cm]
& $\alpha$ &$[0.,1.]$   &$0.29{\pm 0.24}$ \\[0.1cm]
& $w_{0}$ &$[-1.,0.]$   &$-0.61_{-0.13}^{+0.14}$ \\[0.1cm]
& $w_{1}$ &$[0.,1.]$   &$0.225_{- 0.026}^{+0.023}$ \\[0.1cm]
& $w_{2}$ &$[0.,1.]$   &$0.17{\pm 0.13}$ \\[0.1cm]
& $m$  &$[0.,1.]$   & $0.24{\pm 0.18}$  \\[0.1cm]
& $n$  &$[0.,1.]$   & $0.32{\pm 0.10}$  \\[0.1cm]
& $\nu$  &$[0.,2.]$   & $0.741{\pm 0.024}$  \\[0.1cm]
& $\lambda$  &$[0.,2.]$   & $0.930{\pm 0.019}$  \\[0.1cm]
&$r_{d}$ (Mpc)   &$[100,300]$  &$146.8{\pm 2.2}$  \\[0.1cm]
\hline
\end{tabular}
\caption{MCMC Results}\label{tab_1}
\end{table}
\section{Observational and theoretical comparisons of the Hubble function and Hubble difference}\label{sec6}
Comparing CDMMA parametrizations in EG and HL gravity with the widely accepted $\Lambda$CDM model is essential for understanding deviations and gaining insights into the Universe's dynamics. This analysis, using dataset like cosmic chronometers (CC), helps identify differences in expansion rate and overall behavior. By assessing how well our models fit observational data and comparing them with $\Lambda$CDM, we can evaluate their viability and reliability, shedding light on their strengths and limitations. Figure ~\ref{fig_3} illustrates this comparative study, offering valuable insights into the cosmological implications of our parametrizations and their impact on our understanding of the Universe.\\\\
\subsection{Comparison with the Hubble data points}
Fig.~\ref{fig_3a}, illustrate the evolution of the Hubble rate \(H(z)\) as a function of redshift (\(z\)) for two distinct models: the \(\Lambda\)CDM model and CDMMA parameterization in the context of EG and HL gravity frameworks. We assess the compatibility of the CDMMA parameterization within the EG and HL frameworks, represented by the red and pink lines, respectively, with a dataset comprising 31 CC measurements, corresponding with their error bars. For comparative analysis, we include the established \(\Lambda\)CDM paradigm, characterized by (\(\Omega_{\mathrm{m0}} = 0.31\) and \(\Omega_{d0} = 0.68\)), indicated by a black line. At lower redshifts, a slight deviation becomes apparent between the \(\Lambda\)CDM model and the CDMMA parameterization within the context of EG and HL Gravity. As the redshift decreases, this deviation diminishes. Notably, closer inspection reveals a clear alignment emerging between the \(\Lambda\)CDM model and the CDMMA parameterization within the EG and HL Gravity frameworks.  he agreement between our model and the Hubble data supports their viability and highlights their potential to offer meaningful insights into the dynamics and evolution of the Universe. These findings underscore the satisfactory alignment of our parametrizations with the observational data, validating their effectiveness in describing cosmic expansion.
\subsection{Hubble difference between model and $\Lambda$CDM}
\begin{figure*}[htb]
\begin{subfigure}{.42\textwidth}
\includegraphics[width=\linewidth]{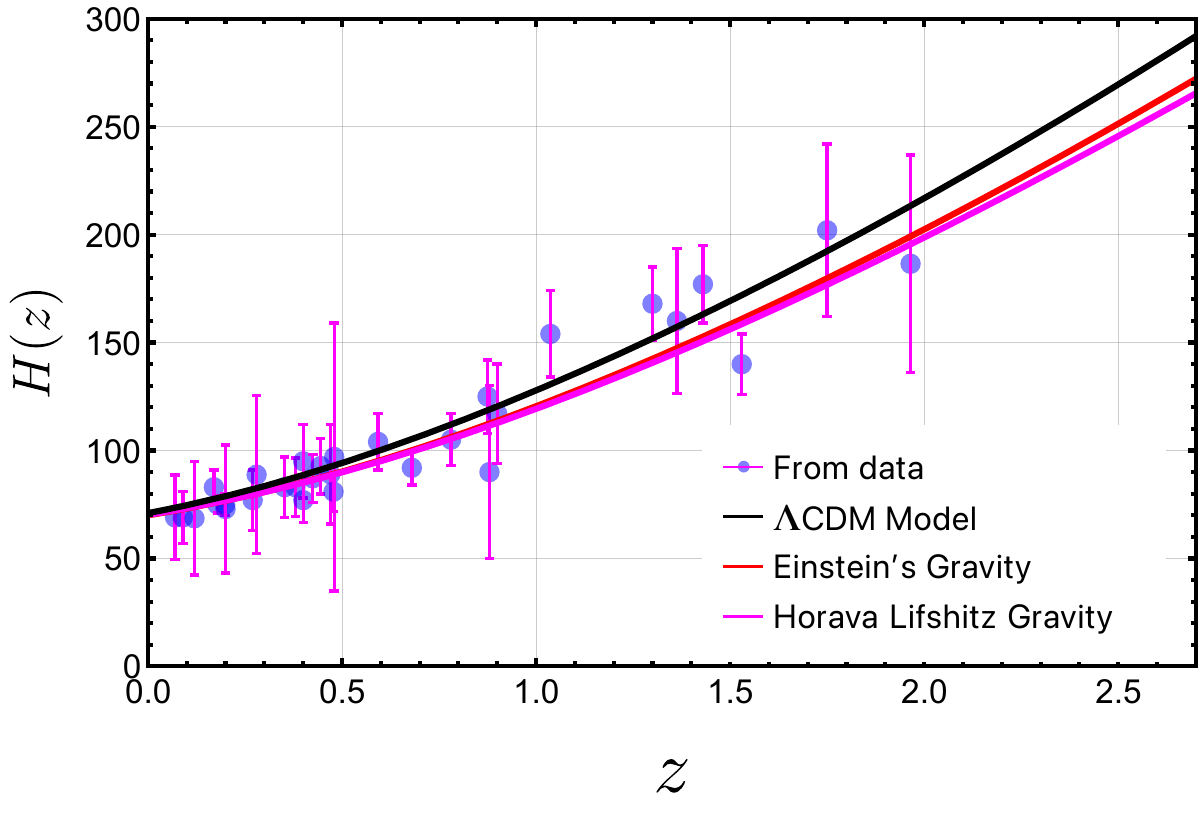}
    \caption{Hubble Parameter $H(z)$}
    \label{fig_3a}
\end{subfigure}
\hfil
\begin{subfigure}{.42\textwidth}
\includegraphics[width=\linewidth]{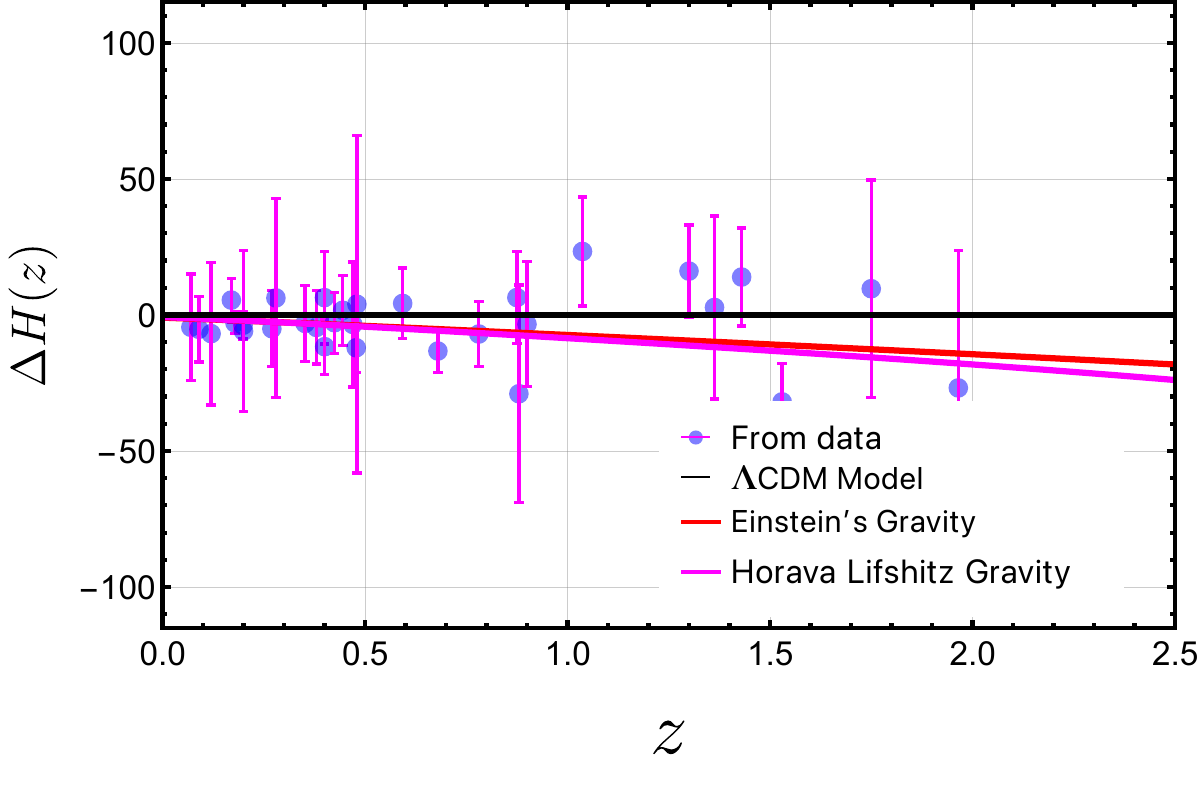}
    \caption{Hubble Difference $\Delta H(z)$}
    \label{fig_3b}
\end{subfigure}
\caption{The figure shows the evolution of the Hubble parameter and Hubble difference versus redshift for the \(\Lambda\)CDM Model (black line) and the CDMMA parameterization in EG and HL Gravity (red and magenta lines, respectively). The data points (purple dots) represent the CC dataset, with error bars shown in magenta.}\label{fig_3}
\end{figure*}

Fig~\ref{fig_3b} illustrates the variation in the difference between the CDMMA parameterization in both frameworks and the $\Lambda$CDM model as a function of redshift (\(z\)). For \(z > 0.5\), a distinct deviation of the CDMMA parameterization in both frameworks becomes evident against the cosmic chronometer (CC) measurements. However, for \(z < 0.5\), this deviation diminishes, and the CDMMA parameterization gradually becomes consistent with the $\Lambda$CDM model. 
\section{Analyzing some Cosmographic Parameters}\label{sec7}
Cosmographic parameters \cite{visser2005cosmography} are crucial in cosmology for describing the universe's large-scale structure and dynamics without relying on specific models. The Hubble Parameter ($H_{0}$) measures the current rate of expansion of the universe, defining the relationship between the recessional velocity of galaxies and their distance from us. The Deceleration Parameter ($q_{0}$) indicates whether the universe's expansion is accelerating or decelerating, with a negative value suggesting acceleration and a positive value indicating deceleration. The Jerk Parameter ($j_{0}$) describes the rate of change of the deceleration parameter, providing insights into the dynamics of the universe's expansion beyond simple acceleration or deceleration. Finally, the Snap Parameter ($s_{0}$) offers higher-order information about the change in the jerk parameter, allowing for a deeper understanding of cosmic acceleration dynamics. These parameters, derived from the Taylor expansion of the scale factor, help analyze observational data and lead to a robust understanding of cosmic evolution.
\subsection{The deceleration parameter}
The deceleration parameter ($q$) \cite{visser2005cosmography}, initially introduced by Edwin Hubble, is a vital cosmological metric for understanding the dynamics of the Universe's expansion. Defined as $q = -\frac{a\ddot{a}}{\dot{a}^2}$, where $a(t)$ represents the scale factor of the Universe over time, it offers crucial insights into cosmic evolution. A positive $q$ indicates a slowing expansion, suggesting past domination by gravitational forces. A $q$ of zero suggests a constant expansion rate, termed a "critical Universe," while a negative $q$ implies accelerating expansion, attributed to dark energy. In modern cosmology, $q$ is pivotal for studying dark energy, dark matter, and overall Universe geometry, serving as a key tool in observational cosmology.
\subsection{The jerk parameter}
The significance of the jerk parameter \cite{visser2004jerk} in cosmology extends beyond traditional parameters like the scale factor and deceleration parameter. It characterizes cosmic dynamics uniquely. We can express the expansion of the scale factor around a reference time using a Taylor series. The jerk parameter, denoted as \(j\), captures information about cosmic evolution and aids in distinguishing between different dark energy proposals. It's a key link between dark energy and conventional universe models. Different values of \(j\) establish connections between various dark energy hypotheses and the standard \(\Lambda\)CDM model, for instance, \(j = 1\) corresponds to the flat \(\Lambda\)CDM model. Understanding '\(j\)' is crucial for exploring cosmic expansion dynamics and transitions between different accelerated expansion eras.
\subsection{Snap parameter}
The snap parameter \cite{visser2004jerk}, denoted as '\( s \)', is a cosmological metric that characterizes the fifth time derivative of the expansion factor in cosmology, offering insights into the curvature and expansion dynamics of the Universe. It is a key component of the Taylor expansion, which describes the Universe's growth. In this expansion, \( s_0 \) represents the fourth-order term, alongside parameters such as \( q_0 \) and \( j_0 \). In the standard \(\Lambda\)CDM model, where \( j = 1 \), the snap parameter simplifies to \( s = -(2 + 3q) \), indicating deviations from this model's expectations. Analyzing \( \frac{ds}{dq} \) relative to \(-3\) provides further insight into these deviations.
\begin{figure*}[htb]
\begin{subfigure}{.32\textwidth}
\includegraphics[width=\linewidth]{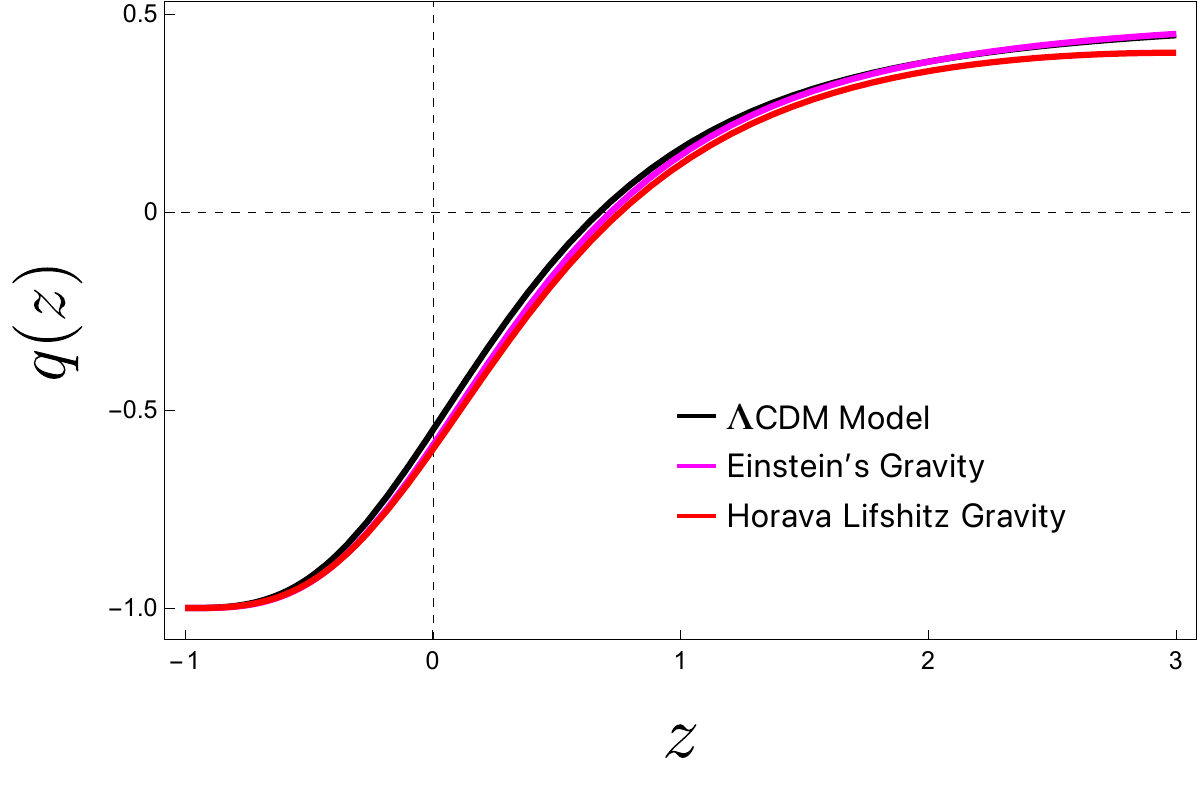}
    \caption{deceleration parameter}
    \label{fig_4a}
\end{subfigure}
\hfil
\begin{subfigure}{.32\textwidth}
\includegraphics[width=\linewidth]{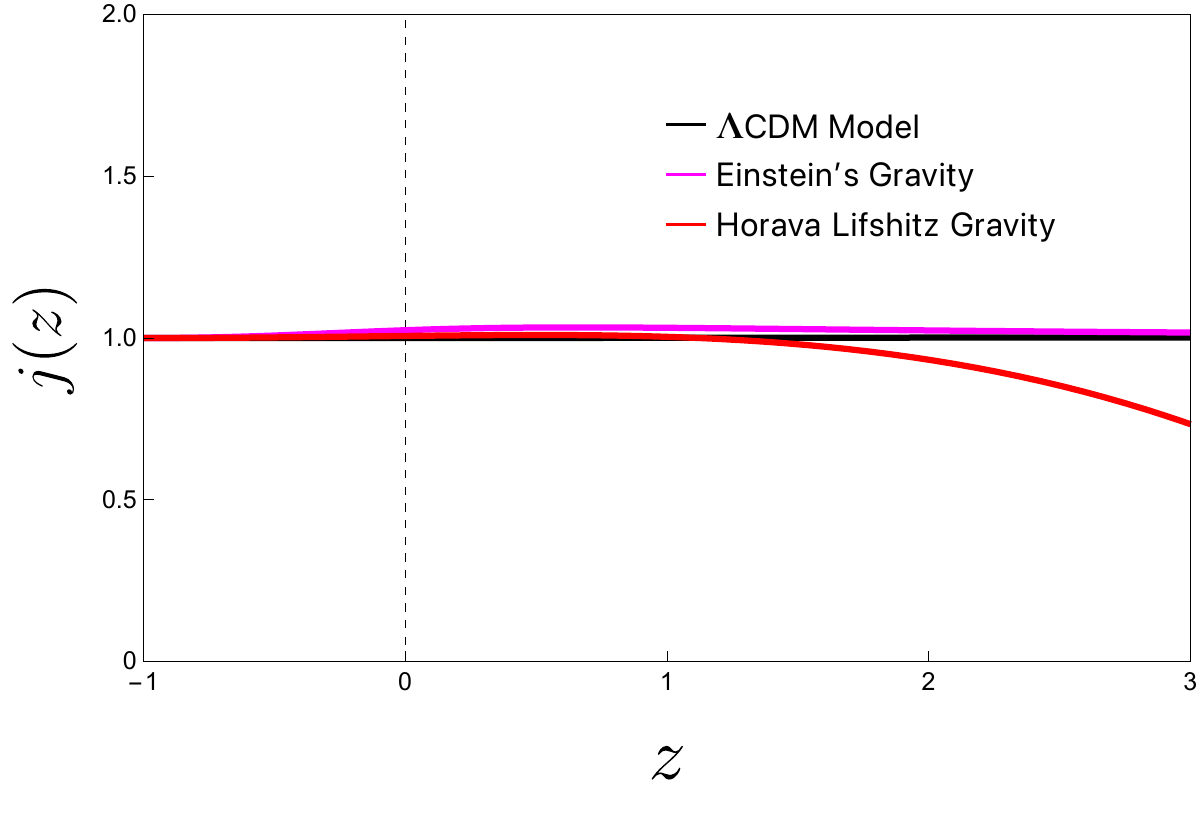}
    \caption{Jerk Parameter}
    \label{fig_4b}
\end{subfigure}
\hfil
\begin{subfigure}{.32\textwidth}
\includegraphics[width=\linewidth]{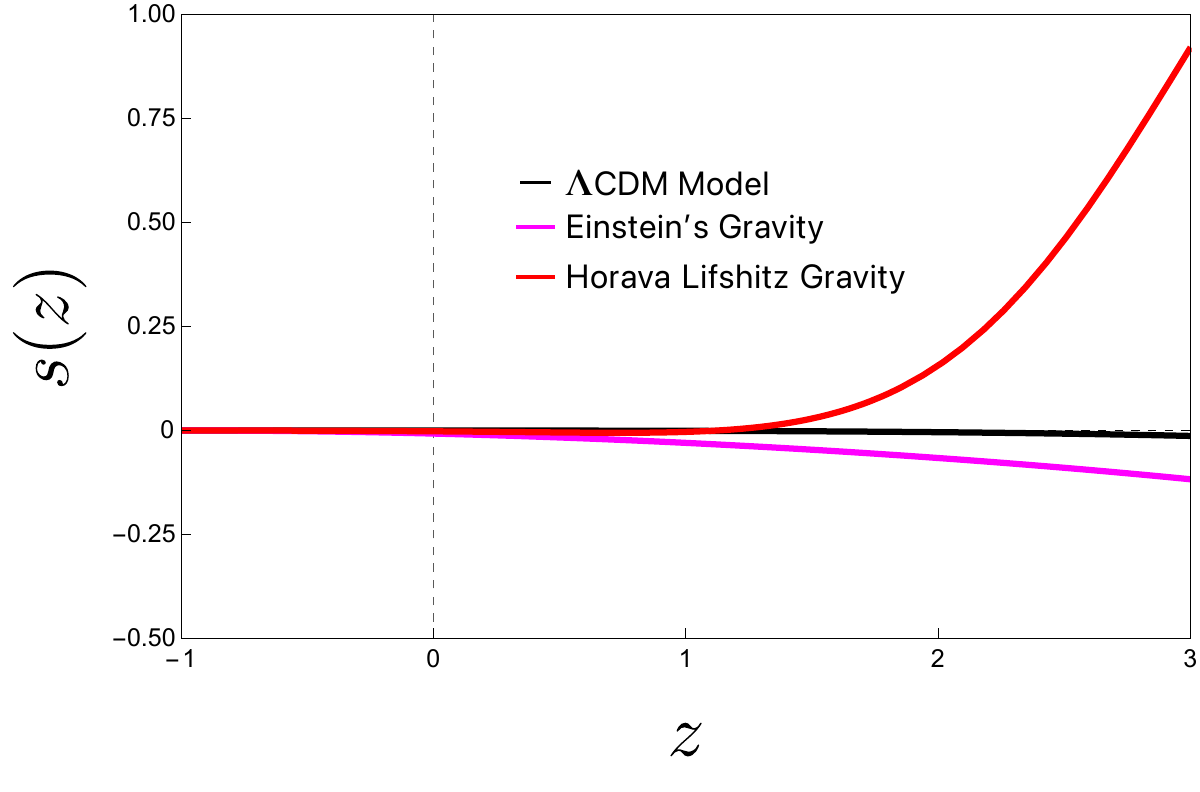}
    \caption{Snap Parameter}
     \label{fig_4c}
\end{subfigure}
\caption{The figure depicts the evolution of various Cosmography parameters with respect to redshift ($z$) is depicted. The red line represents the $\Lambda$CDM model, while the magenta line represents the CDMMA parameterization within EG and HL Gravity. These lines utilize the best-fit values obtained from the Joint analysis.}\label{fig_4}
\end{figure*}
\section{Diagnostic Analysis of the models}\label{sec8}
\subsection{Statefinder diagnostic}
\cite{sahni2003statefinder,alam2003exploring,alam2003exploring} proposed a novel method for diagnosing dark energy (DE) based on higher derivatives of the scale factor. Known as statefinder diagnostics, this technique contrasts different DE models by utilizing higher-order derivatives of the scale factor. The diagnostic pair \(\{r,s\}\) allows for a model-independent examination of DE cosmic properties, defined by the relationships \(r=\frac{\dddot{a}}{aH^{3}}\) and \(s=\frac{r-1}{3\left( q-\frac{1}{2}\right)}\). Parameter \(s\) is a linear combination of \(r\) and \(q\), offering dimensionless and geometric insights. Different regions in the \(\{r,s\}\) and \(\{q,r\}\) planes represent various DE models, with specific pairs correlating to classic models such as \(\{r,s\}=\{1,0\}\) for the \(\Lambda\)CDM model and \(\{r,s\}=\{1,1\}\) for standard cold dark matter (SCDM). Trajectories in the \(r-s\) plane delineate quintessence-like and phantom-like DE models, where \(s>0\) and \(s<0\) respectively. Deviations from standard values (\(\{r,s\}={1,0}\)) signify a transition between phantom and quintessence models. Similarly, specific \(\{q,r\}\) pairs correspond to standard DE models like \(\{-1,1\}\) for \(\Lambda\mathrm{CDM}\) and \(\{0.5,1\}\) for SCDM. Deviations from these standard values indicate non-conventional cosmic models.
\subsection{$O_{m}$ Diagnostic}
In our research, we employ the \( O_{m} \) diagnostic for Dark Energy (DE), as introduced by \cite{sahni2008two}. This diagnostic is notable for its simplicity and reliance on the directly observable Hubble parameter \( H(z) \). The \( O_{m} \) diagnostic effectively distinguishes between different cosmological models, specifically the cosmological constant (\(\Lambda\)CDM) and dynamic models (curved \(\Lambda\)CDM), using \( O_{m} \) and \(\Omega_{m0}\) as priors. If \( O_{m} = \Omega_{m0} \), it aligns with the \(\Lambda\)CDM model. Conversely, \( O_{m} > \Omega_{m0} \) indicates a quintessence scenario, while \( O_{m} < \Omega_{m0} \) suggests a phantom scenario \cite{escamilla2016nonparametric}. This diagnostic is a powerful tool for distinguishing between cosmological models due to its reliance on observational data, particularly the Hubble parameter. In a flat Universe, \( O_{m} \) is defined as:
\begin{equation}
O_{m} = \frac{\left( \frac{H(z)}{H_{0}} \right)^{2} - 1}{(1+z)^{3} - 1}.
\end{equation}
\begin{figure*}[htb]
\begin{subfigure}{.32\textwidth}
\includegraphics[width=\linewidth]{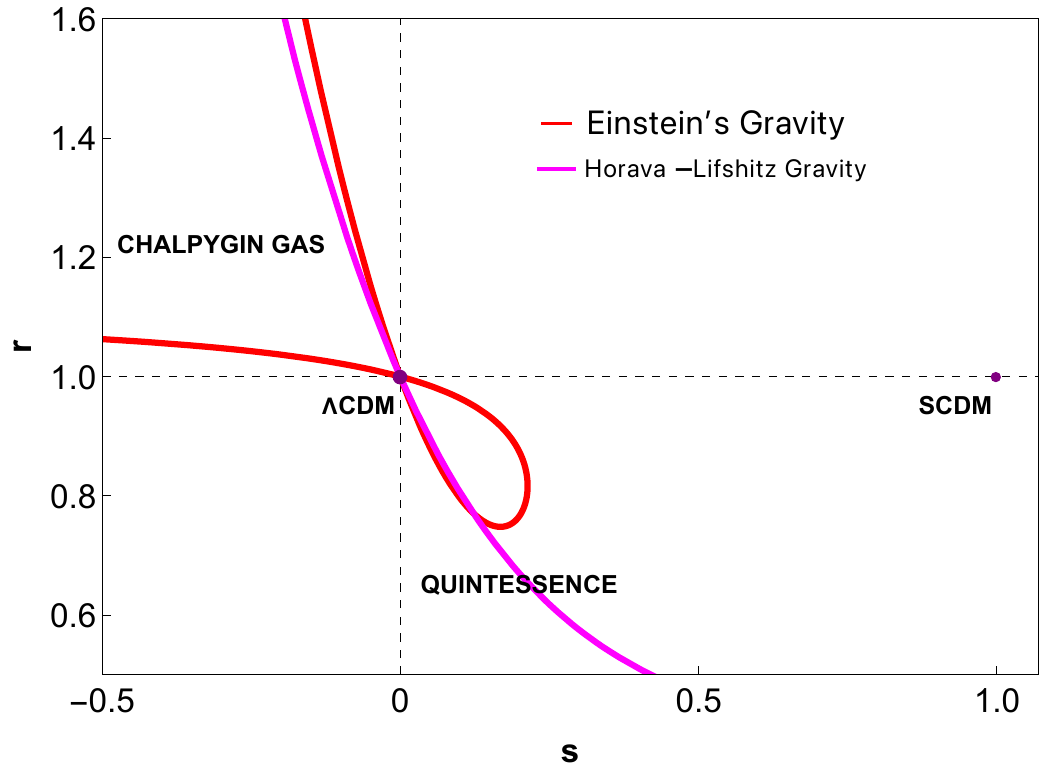}
    \caption{$\{s, r\}$ profile}
    \label{fig_6a}
\end{subfigure}
\hfil
\begin{subfigure}{.32\textwidth}
\includegraphics[width=\linewidth]{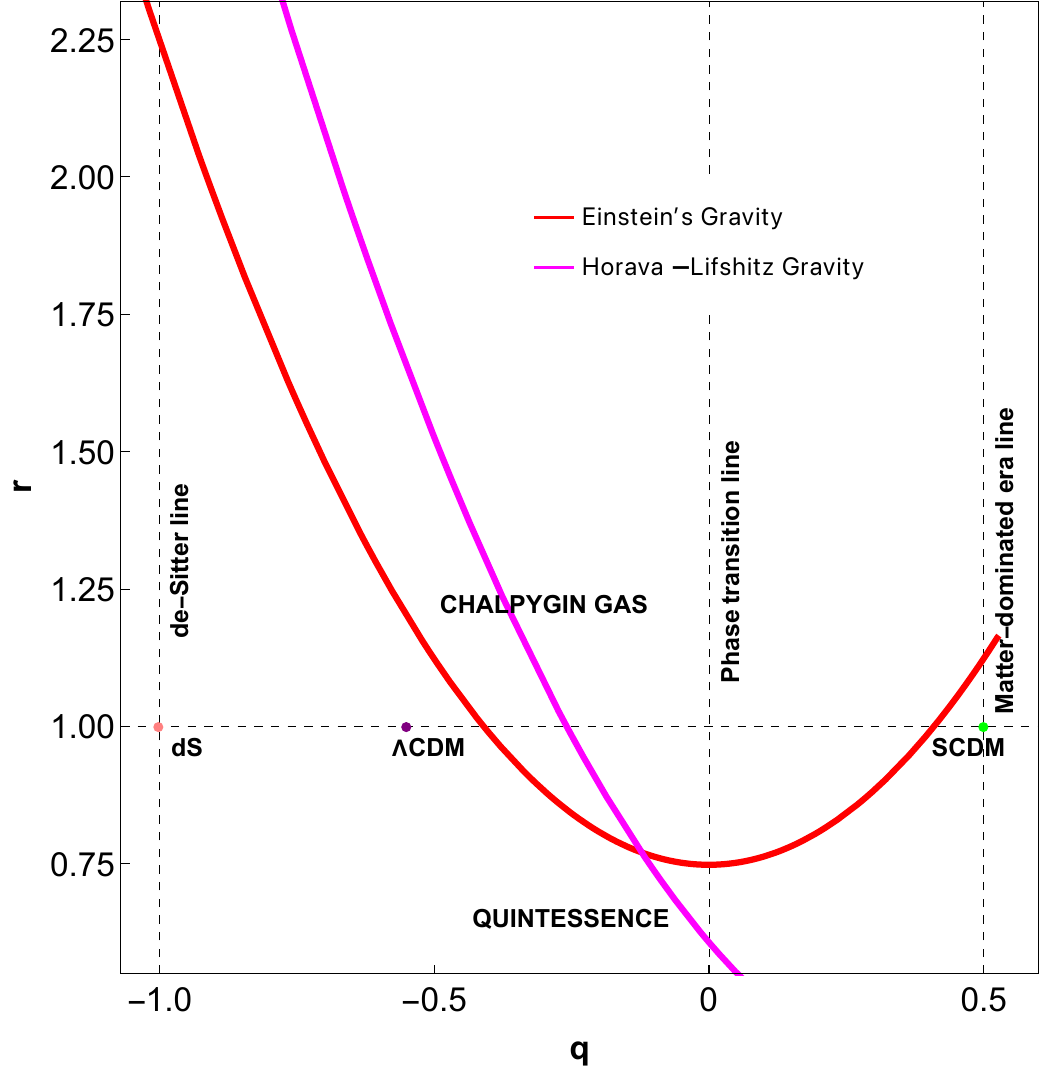}
    \caption{$\{q, r\}$ profile}
    \label{fig_6b}
\end{subfigure}
\hfil
\begin{subfigure}{.32\textwidth}
\includegraphics[width=\linewidth]{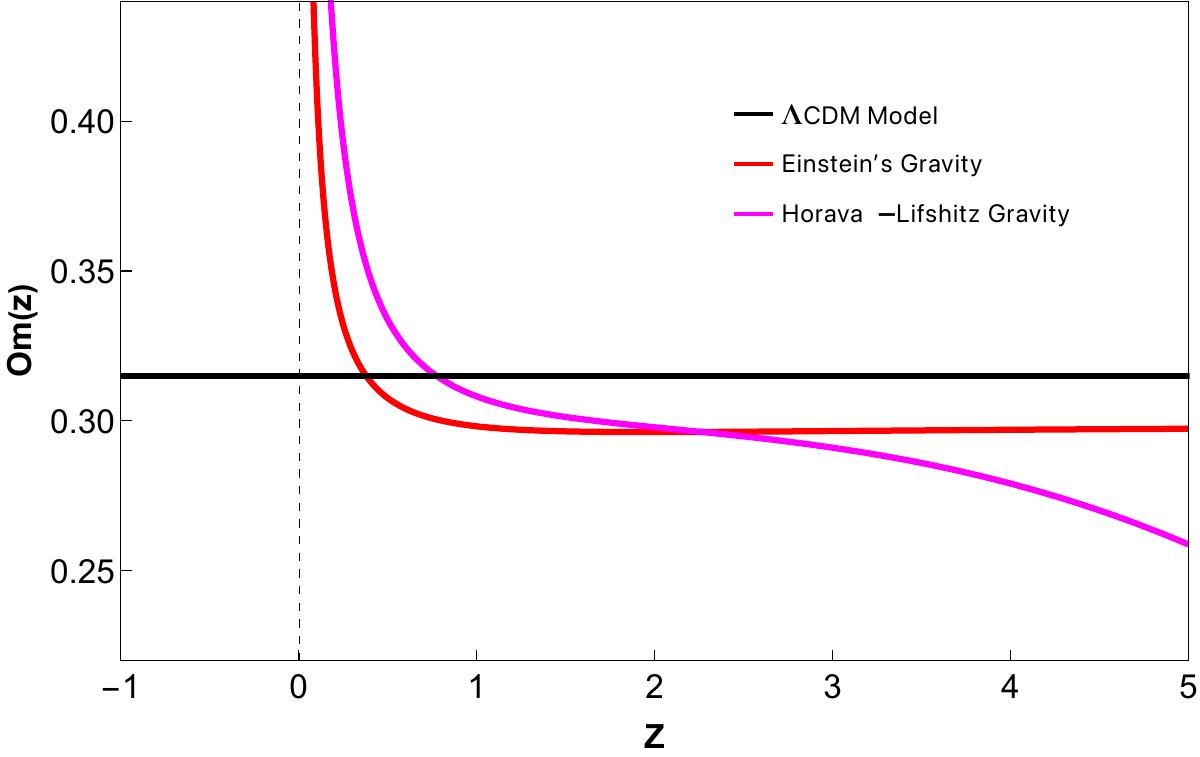}
    \caption{$O_{m}$ Diagnostic}
     \label{fig_6c}
\end{subfigure}
\caption{The evolution of the Statefinder diagnostic and the $O_{m}$ diagnostic profile in the CDMMA parameterization within EG and HL Gravity is depicted by the red and magenta lines, respectively. These lines utilize the best-fit values obtained from the joint analysis.}\label{fig_6}
\end{figure*}
\section{Statistical Analysis}\label{sec4}
To assess the performance of the CDMMA parameterization model, we apply two widely used model selection criteria: the Akaike Information Criterion (AIC) and the Bayesian Information Criterion (BIC) \cite{AIC1,AIC2,AIC3,AIC4,AIC,BIC}. These metrics evaluate the trade-off between model fit and complexity. The AIC is defined as: $\text{AIC} = \chi_{\text{tot},\min}^2 + 2k + \frac{2k(k + 1)}{N_{\text{tot}} - k - 1},$
where \(\chi_{\text{tot},\min}^2\) represents the minimized total chi-squared, \(N_{\text{tot}}\) is the total number of data points, and \(k\) is the number of model parameters. For large datasets where \(N_{\text{tot}}\) is large, AIC simplifies to its standard form: $\text{AIC} \approx \chi_{\text{tot},\min}^2 + 2k,$
balancing the goodness of fit (via \(\chi_{\text{tot},\min}^2\)) and penalizing model complexity through \(2k\). The BIC is given by: $\text{BIC} = \chi_{\text{tot},\min}^2 + k \ln(N_{\text{tot}}),$
which penalizes model complexity more strongly than AIC for large datasets, reflecting the stricter nature of BIC in model selection. The reduced chi-squared statistic \cite{andrae2010and} is another key metric that evaluates the overall fit quality of the model: $\chi_{\text{red}}^2 = \frac{\chi_{\text{tot},\min}^2}{N_{\text{tot}} - k},$
where \(\chi_{\text{red}}^2 \approx 1\) suggests a good fit, as it implies that the model explains the data within expected statistical fluctuations. A value of \(\chi_{\text{red}}^2 < 1\) might indicate overfitting, where the model is too complex. Conversely, \(\chi_{\text{red}}^2 \gg 1\) suggests a poor fit, indicating that the model does not capture important data features. To compare models, we compute the differences \(\Delta \text{AIC}\) and \(\Delta \text{BIC}\) relative to a baseline model, typically the \(\Lambda\)CDM model. Negative values of \(\Delta \text{AIC}\) or \(\Delta \text{BIC}\) indicate that the model is preferred over the baseline. According to \cite{jeffreys1998theory}, if \(0 < |\Delta \text{AIC}| \leq 2\), the models are considered comparable, whereas \( |\Delta \text{AIC}| \geq 4 \) indicates the model with the higher AIC is significantly less favored. For \(\Delta \text{BIC}\), if \(0 < |\Delta \text{BIC}| \leq 2\), disfavor is weak, \(2 < |\Delta \text{BIC}| \leq 6\) indicates strong disfavor, and \( |\Delta \text{BIC}| > 6\) points to very strong disfavor.
\begin{table*}[htbp]
\begin{center}
\begin{tabular}{|c|c|c|c|c|c|c|c|c|}
\hline
Model & ${\chi_{\text{tot},min}^2}$ & $N_{tot}$ & $k$ & $\chi_{\text {red}}^2$ & AIC & $\Delta$AIC & BIC & $\Delta$BIC \\[0.1cm]
\hline
$\Lambda$CDM Model & 1801.01 & 1758 & 4 & 1.026 & 1807.01 & 0 & 1823.42 & 0 \\[0.1cm] \hline
\text{Model 1} & 1758.31 & 1758 & 11 & 1.006 & 1780.31 & -26.70 & 1840.50 & 17.08 \\[0.1cm] \hline
\text{Model 2} & 1767.12 & 1758 & 13 & 1.013 & 1793.12 & -13.89 & 1864.26 & 40.84 \\[0.1cm] \hline
\end{tabular}
\caption{Summary of ${\chi_{\text{tot},min}^2} $, $\chi_{\text {red }}^2$, AIC, $\Delta$AIC, BIC, $\Delta$BIC for the $\Lambda$CDM Model and CDMMA parameterization within EG and HL Gravity}\label{tab_2}
\end{center}
\end{table*}
\section{Results}\label{sec9}
In our analysis, we use BAO measurements from DESI Year 1 and SDSS-IV, the Pantheon$^{+}$ sample, and cosmic chronometer (CC) measurements to compare the $\Lambda$CDM model and the CDMMA parameterization within both EG and HL gravity frameworks. We examine the parameters \( H_0 \) (km s\(^{-1}\) Mpc\(^{-1}\)), \( \Omega_{m0} \), \( \Omega_{k0} \), \( \Omega_{d0} \), and \( r_d \) (Mpc). In the $\Lambda$CDM model, we find \( H_0 = 68.2 \pm 1.1 \, \text{km} \, \text{s}^{-1} \, \text{Mpc}^{-1} \), consistent with values reported by \cite{aghanim2020planck}. The CDMMA parameterization, however, yields \( H_0 = 69.9 \pm 1.6 \, \text{km} \, \text{s}^{-1} \, \text{Mpc}^{-1} \) in EG and \( H_0 = 66.3 \pm 1.0 \, \text{km} \, \text{s}^{-1} \, \text{Mpc}^{-1} \) in HL gravity, aligning with values reported by \cite{aghanim2020planck}. For the matter density parameter, the $\Lambda$CDM model predicts \( \Omega_{m0} = 0.311 \), while the CDMMA parameterization predicts \( \Omega_{m0} = 0.2877 \) in EG and \( \Omega_{m0} = 0.2832 \) in HL gravity, showing close agreement with recent estimates from the DESI collaboration. The spatial curvature parameter, \( \Omega_{k0} \), is approximately \( 0.001 \) in EG and \( -0.001\) in HL gravity in CDMMA parameterization, suggesting consistency with a flat Universe in both frameworks, as previously predicted by \cite{hinshaw2013nine}. For the dark energy density parameter, the $\Lambda$CDM model predicts \( \Omega_{d0} = 0.686 \), while the CDMMA parameterization gives \( \Omega_{d0} = 0.698 \) in EG and \( \Omega_{d0} = 0.694 \) in HL gravity, consistent with estimates from Planck 2018 \cite{aghanim2020planck}. The sound horizon at the drag epoch, \( r_d \), is a crucial parameter in cosmology. The $\Lambda$CDM model predicts \( r_d = 146.6 \pm 2.4\, \text{Mpc} \), while the CDMMA parameterization yields \( r_d = 146.6 \pm 2.4\, \text{Mpc} \) in EG and \( r_d = 146.8 \pm 2.2\, \text{Mpc} \) in HL gravity, values closely aligned with \cite{aghanim2020planck}. Fig. \ref{fig_4} illustrates the evolution of the deceleration parameter (DP) \( q \), the jerk parameter (JP) \( j \), and the snap parameter \( s \) for the $\Lambda$CDM Model and CDMMA parameterization in EG and HL gravity. Fig. \ref{fig_4a} represents the evolution of DP at various epochs. In the $\Lambda$CDM model, the deceleration parameter \( q \) exhibits distinctive values at different cosmic epochs. At high redshifts (\( z \rightarrow \infty \)), the initial deceleration parameter \( q_i \) is measured at 0.449061. As the Universe evolves to its present state (\( z \rightarrow 0 \)), \( q_0 \) reaches -0.550, indicating a transition from decelerated to accelerated expansion. In the distant future (\( z \rightarrow -1 \)), \( q_f \) approaches -1, suggesting a Universe dominated by dark energy, leading to accelerated expansion. The transition redshift (\( z_{tr} \)), where \( q = 0 \), occurs at 0.682. Comparatively, the CDMMA parameterization in EG and HL gravity exhibit similar evolutionary trends in the deceleration parameter \( q \) across cosmic epochs. At high redshifts, the CDMMA parameterization shows comparable \( q_i \) values: 0.450 in EG and 0.390723 in HL gravity. Presently, \( q_0 \) is slightly lower in EG (-0.599) compared to HL gravity (-0.601). However, the transition redshift \( z_{tr} \) in EG (0.711) is higher than that in HL gravity (0.769), indicating differences in the rate of cosmic acceleration between the two different frameworks. In the distant future (\( z \rightarrow -1 \)), in both frameworks, the DP approaches -1, suggesting a universe dominated by dark energy, leading to accelerated expansion. Fig. \ref{fig_4b} depicts the evolution of the jerk parameter (JP). At \(z < 1.3\), the jerk parameter remains constant with a value of 1 in both the EG and HL gravity frameworks. However, for \(z > 1.3\), the values of the jerk parameters for the CDMMA parameterization in HL gravity deviate from those obtained in both $\Lambda$CDM and the CDMMA parameterization in EG Gravity. The consistent values of the CDMMA parameterization and $\Lambda$CDM at late times imply that the rate of change of acceleration with respect to time remains constant throughout the evolution of the Universe. Fig. \ref{fig_4c} illustrates the evolution of the snap parameter (SP), offering insights into the cosmic expansion dynamics across epochs for the CDMMA parameterization in EG and HL gravity compared to the standard $\Lambda$CDM model. In the $\Lambda$CDM model, at high redshifts (\( z \rightarrow \infty \)), \( s_i \) takes a substantial value of 3.359387. As the Universe evolves to the present (\( z \rightarrow 0 \)), \( s_0 \) decreases to 0.357323. In the distant future (\( z \rightarrow -1 \)), \( s_f \) approaches -1, suggesting a reversal of cosmic acceleration. Comparatively, in the CDMMA parameterization in EG, the snap parameter behaves similarly to the $\Lambda$CDM model across epochs. However, for \(z > 0.37\), the values of the snap parameters for the CDMMA parameterization in HL gravity deviate from those obtained in both $\Lambda$CDM and the CDMMA parameterization in EG. In Fig. \ref{fig_6a}, we can observe how the $\{s, r\}$ profile of the CDMMA parameterization behaves in both EG and HL gravity. Initially, in EG, the model exhibits values where $r > 1$ and $s < 0$, indicating the quintessence region. As the evolution progresses, the model transitions, crossing the fixed point at $\{r, s\}=\{1, 0\}$, which represents the $\Lambda$CDM model. It then takes the values of $r < 1$ and $s > 0$, corresponding to the Chaplygin gas region. Subsequently, the model crosses the fixed point again at $\{r, s\}=\{1, 0\}$, returning to the quintessence region with $r > 1$ and $s < 0$. In contrast, in HL gravity, the CDMMA parameterization initially exhibits values in the quintessence region with $r > 1$ and $s < 0$. However, as evolution progresses, it crosses the fixed point at $\{r, s\}=\{1, 0\}$, transitioning to values where $r < 1$ and $s > 0$, corresponding to the Chaplygin gas region. Fig \ref{fig_6b} illustrates how the ${q, r}$ profile of the CDMMA parameterization behaves in both EG and HL gravity. Initially, in the EG frame, the model manifests values within the range $q > 0$ and $r > 1$, characterized as the matter-dominated phase. As the evolution progresses, the model undergoes a phase transition, leading to values within the range $q < 0$ and $r < 1$, indicative of the quintessence region. As the model evolves, it shifts towards values within the range $q < 0$ and $r > 1$, corresponding to the Chaplygin gas region. The negative $q$ signifies an accelerating expansion of the Universe, while $r > 1$ implies dominance of the Chaplygin gas over matter, driving the acceleration. Ultimately, the model converges towards the de Sitter line situated at $q = -1$. In the case of HL gravity, the model exhibits a similar behavior but at a slower rate (Initially $q > 0$, $r > 1$ (matter), transitioning to $q < 0$, $r < 1$ (quintessence), then $q < 0$, $r > 1$ (Chaplygin gas), converging to $q = -1$). Fig \ref{fig_6c} illustrates the evolution of the $O_{m}$ diagnostic of the CDMMA parameterization in both EG and HL gravity . Both models exhibit similar behavior: at high redshifts, the value of $O_{m}$ consistently remains below the current matter density parameter $\Omega_{m0}$, suggesting that the models reside in the phantom region. However, as evolution progresses, $O_{m}$ becomes higher than $\Omega_{m0}$, indicating that both models transition into the quintessence region. Table \ref{tab_2} presents a comparative analysis of the $\Lambda$CDM Model, Model 1, and Model 2 based on various statistical metrics. The $\Lambda$CDM Model has the highest total chi-squared value (${ \chi_{\text{tot},min}^2}$) of 1801.01, suggesting it does not fit the data as well as the alternative models. Model 1 improves upon this with a lower total chi-squared of 1758.31, indicating a better fit. Model 2 also offers a better fit than $\Lambda$CDM, with a value of 1767.12, though not as good as Model 1. The reduced chi-squared values, which account for the number of data points and parameters, show a similar pattern: Model 1 has the lowest value (1.006), slightly better than $\Lambda$CDM (1.026) and Model 2 (1.013). In terms of model efficiency, the Akaike Information Criterion (AIC) for the $\Lambda$CDM Model is the highest at 1807.01, suggesting it is the least efficient. Model 1, with an AIC of 1780.31, is more efficient, while Model 2, with an AIC of 1793.12, falls between the two. The $\Delta$AIC, which indicates improvement over the $\Lambda$CDM Model, is the largest for Model 1 (-26.70), followed by Model 2 (-13.89). The Bayesian Information Criterion (BIC) follows a similar trend, with $\Lambda$CDM having the highest BIC (1823.42), Model 1 having a lower BIC (1840.50), and Model 2 having the highest BIC among the models (1864.26). The $\Delta$BIC also shows that Model 1 improves upon the $\Lambda$CDM Model more effectively than Model 2. In conclusion, while Model 2 provides a better fit than the $\Lambda$CDM Model, Model 1 offers the best balance between fit and efficiency, making it the most favorable model in this comparison.

\section{Summary and Conclusions}\label{sec10}
In this manuscript, we explored the parameter constraints and conducted a comparative analysis of the standard \(\Lambda\)CDM model and the CDMMA parameterization within the frameworks of EG and HL gravity. Our study utilized multiple observational datasets, including cosmic chronometer, Type Ia Supernovae, and Baryon Acoustic Oscillation (BAO) datasets from DESI-Y1 and SDSS-IV. Figures \ref{fig_1} and \ref{fig_2} illustrate the parameter constraints with 1\(\sigma\) and 2\(\sigma\) confidence levels for the CDMMA parameterization in both EG and HL gravity. Table \ref{tab_1} presents the best-fit values of cosmological parameters for the \(\Lambda\)CDM model, as well as for the CDMMA parameterization in EG and HL gravity. Fig.~\ref{fig_3} compared the \(\Lambda\)CDM model and the CDMMA parameterization within EG and HL gravity frameworks using dataset such as CC. The evolution of the Hubble rate \(H(z)\) as a function of redshift was presented in Fig.~\ref{fig_3a}, showing a slight deviation between the \(\Lambda\)CDM model and the CDMMA parameterization at lower redshifts, which diminishes as redshift decreases. Fig.~\ref{fig_3b} depicted the difference between the CDMMA parameterization and the \(\Lambda\)CDM model, revealing a distinct deviation at higher redshifts that diminishes over time. Fig.~\ref{fig_4} illustrated the evolution of the deceleration parameter (DP), jerk parameter (JP), and snap parameter (SP) for both the \(\Lambda\)CDM model and the CDMMA parameterization in EG and HL gravity. The deceleration parameter exhibited similar evolutionary trends across cosmic epochs for CDMMA parameterization in both frameworks, with slight differences in the transition redshift and present-day values. The jerk and snap parameters highlighted differences at high redshifts, particularly in the HL gravity framework, which deviated from the \(\Lambda\)CDM model. Fig.~\ref{fig_6a} and Fig.~\ref{fig_6b} illustrated the \(\{s, r\}\) and \(\{q, r\}\) profiles of the CDMMA parameterization, respectively. The \(\{s, r\}\) profile showed transitions through the quintessence and Chaplygin gas regions, with the CDMMA parameterization eventually aligning with the \(\Lambda\)CDM model. The \(\{q, r\}\) profile demonstrated a similar behavior, with phase transitions indicating the cosmic acceleration and eventual convergence towards a de Sitter Universe. Fig.~\ref{fig_6c} depicted the evolution of the \(O_{m}\) diagnostic, showing a transition from the phantom region to the quintessence region for both EG and HL gravity. In our model comparison, we used the Akaike Information Criterion (AIC) and the Bayesian Information Criterion (BIC) to evaluate the models. Although the \(\Lambda\)CDM model showed the strongest fit in terms of total chi-squared (${ \chi_{\text{tot},min}^2}$), the AIC and BIC values favored the CDMMA parameterization, particularly Model 1. Model 1 demonstrated a significantly lower AIC (1780.31) compared to the \(\Lambda\)CDM model (1807.01) and a smaller $\Delta$AIC of -26.70, indicating higher efficiency. Model 2 also performed better than \(\Lambda\)CDM, but with a smaller AIC reduction and a higher BIC. The reduced chi-squared values for all models were close to 1, showing satisfactory alignment with the data, with \(\Lambda\)CDM at 1.026, Model 1 at 1.006, and Model 2 at 1.013. Overall, our analysis demonstrates that the CDMMA parameterization within both the EG and HL gravity frameworks show good agreement with observational data. The detailed comparison and constraints provided in this section highlight the potential of these models to enhance our understanding of the Universe's expansion history and the dynamics of dark energy.\\\\
\section*{Acknowledgements}
NUM would like to thank  CSIR, Govt. of
India for providing Senior Research Fellowship (No. 08/003(0141))/2020-EMR-I). G. Mustafa is very thankful to Prof. Gao Xianlong from the Department of Physics, Zhejiang Normal University, for his kind support and help during this research. Further, G. Mustafa acknowledges grant No. ZC304022919 to support his Postdoctoral Fellowship at Zhejiang Normal University.
\bibliographystyle{elsarticle-num}
\bibliography{mybib,mybib2}
\end{document}